\documentclass[11pt,a4paper,epsfig]{article}
\usepackage{color}
\usepackage{amsfonts}
\usepackage{epsfig}

\begin{document}

\title{\bf From ${\cal N}=2$ supersymmetric classical \\ to quantum mechanics and back:\\
the SUSY WKB approximation}

\author{M.A. Gonz\'alez Le\'on$^1$, J. Mateos Guilarte$^2$ and M. de la Torre Mayado$^2$\\
\\ {\small \sl $^1$Departamento de Matem\'atica
Aplicada. Universidad de Salamanca, SPAIN.}
\\ {\small \sl $^2$Departamento de F\'{\i}sica Fundamental.
Universidad de Salamanca, SPAIN.}}

\date{}
\maketitle

\begin{abstract}
Links between supersymmetric classical and quantum mechanics are
explored. Diagrammatic representations for $\hbar$-expansions of
norms of ground states are provided. The WKB spectra of
supersymmetric non harmonic oscillators are found.
\end{abstract}

\section{Introduction}

In this essay, written to commemorate the sixtieth birthday of J.
Cari$\tilde{\rm n}$ena, we discuss several elementary issues in
one-dimensional supersymmetric quantum mechanics. The r$\hat{{\rm
o}}$le of the Riccati equation in this framework has been
thoroughly analyzed by Cari$\tilde{\rm n}$ena and collaborators at
the highest level of mathematical rigor by approaching topics such
as the factorization method or shape invariance from a
group-theoretical point of view, see \cite{Car}, \cite{Car1} and
\cite{Car2}. Our purpose here is to approach these matters from a
rather physical point of view. To construct a supersymmetric
quantum mechanical system starting from a physical potential
energy we shall be led to deal with the Hamilton-Jacobi or the
Poisson equations, although in both cases there is an associated
Riccati equation. We shall focus on studying the relationship
between supersymmetric classical and quantum mechanical systems,
following the standard References \cite{Coop} and \cite{Junk} and
the more recent Lectures of A. Wipf \cite{Wipf}. In particular,
models where supersymmetry is unbroken and instantons exist will
be analyzed at length. The main motivation to discuss these 1D
SUSY QM models is to take profit of the knowledge acquired to
study highly non-trivial 2D systems as those proposed in
\cite{Ioffe3}. Another issue to be treated with care is the
semiclassical behavior of supersymmetric quantum systems, this
done with the help of the enlightening paper of A. Comtet et al.
\cite{Com}.

\section{R$\hat{{\rm o}}$le of the Hamilton-Jacobi, Riccati and Poisson \\ equations in
SUSY quantum mechanics}

Let us start with a natural Lagrangian of one degree of freedom
and the action functional:
\begin{equation}
S=\int \, dt
\left[\frac{m}{2}\frac{dx}{dt}\frac{dx}{dt}-V(x;\lambda ,
k)\right] \qquad , \qquad \qquad  [\lambda]={\rm ML}^{-2}{\rm
T}^{-2} \ , \ [k]={\rm MT}^{-2} \qquad . \label{eq:dim}
\end{equation}
We shall consider potential energies $V(x;\lambda, k)$ that depend
on two parameters $\lambda$ and $k$ of dimensions given in
(\ref{eq:dim}) and we shall introduce the non-dimensional
variables: $ x\rightarrow \sqrt{\frac{k}{\lambda}}\cdot x $,
$t\rightarrow \sqrt{\frac{m}{k}} \cdot t$, $V(x;\lambda ,
k)=\frac{k^2}{\lambda}\, V(x)$, such that the action and the
Hamiltonian read (non-dimensional variables will be used in what
follows):
\[
S = \frac{k^{\frac{3}{2}}m^{\frac{1}{2}}}{\lambda}\cdot \int \, dt
\left[\frac{1}{2}\frac{dx}{dt}\frac{dx}{dt}-V(x)\right] \ , \qquad
H(p,x)= \frac{k^2}{\lambda}\,\left({1\over 2}p^2+V(x)\right) \
,\qquad \ p=\frac{\partial L}{\partial \dot{x}}=\frac{dx}{dt}
\qquad .
\]

\subsection{One-dimensional ${\cal N}=2$
SUSY classical mechanics}

A ${\cal N}=2$ supersymmetric extension of a classical mechanical
system of one degree of freedom is constructed as follows:

\noindent {\bf 1.} We add two \lq\lq fermionic" degrees of freedom
to the \lq\lq bosonic" degree of freedom with the real coordinate
$x$. The fermionic coordinates form a Grassman Majorana spinor:
\[
\theta=\left(\begin{array}{c}\theta_1 \\
\theta_2\end{array}\right) \hspace{1.5cm} , \hspace{1.5cm}
\theta_\alpha\theta_\beta+\theta_\beta\theta_\alpha=0 \qquad
,\qquad \forall \, \alpha , \beta=1,2 \, \, \, \qquad .
\]

\noindent {\bf 2.} A superPoisson structure is defined in the
phase superspace with coordinates $p, x, \theta_1, \theta_2$  .
Given two superfunctions $F$ and $G$ on the superspace, the
Poisson superbracket
\[
 \{F,G\}_P=\frac{\partial F}{\partial p} \frac{\partial
G}{\partial x} - \frac{\partial F}{\partial x} \frac{\partial G
}{\partial p}+ i \sum_{\alpha=1}^2 \,F
\frac{\stackrel{\leftarrow}{\partial}}{\partial \theta_\alpha}
\frac{\stackrel{\rightarrow}{\partial}}{\partial \theta_\alpha} G
\]
is read from the basic brackets: $\forall \, \alpha , \beta=1,2 \,
, \, $, $\{p,x\}_P=1$, $\{x,x\}_P=0$, $\{p,p\}_P=0$,
$\{\theta_\alpha, \theta_\beta\}_P=i \delta_{\alpha\beta}$.

\noindent Note that in the \lq\lq soul" of the system - the
subspace of the superspace spanned by the Grassman variables- the
configuration space and the phase space coincide. The reason is
that the Lagrangian ruling the dynamics of the fermionic variables
is of first order in time derivatives. Thus, the time derivatives
of Grassman variables will not appear in the Hamiltonian.

\noindent {\bf 3.} The classical SUSY charges:
$Q_1=p\theta_1-\frac{dW}{dx}\theta_2$,
$Q_2=p\theta_2+\frac{dW}{dx}\theta_1$, close the classical
supersymmetric algebra:
\[
\{ Q_1,Q_1 \}_P=\{ Q_2,Q_2 \}_P=2iH_S \qquad , \qquad \{Q_\alpha ,
H_S\}_P=0 \qquad , \qquad \{ Q_1,Q_2 \}_P=0 \qquad .
\]

\noindent {\bf 4.} The classical Hamiltonian $H_S$
\begin{equation}
H_S=\frac{1}{2} p^2 + \frac{1}{2}\frac{dW}{dx}\frac{dW}{dx} - i
\frac{d^2W}{dx^2} \theta_2 \theta_1  \, \, \label{eq:hams}
\end{equation}
is invariant by construction with respect to the
super-transformations generated by $Q_1$ and $Q_2$. Besides the
kinetic energy of the bosonic variables, there are two interaction
energy terms in the Hamiltonian (\ref{eq:hams}) proportional to
the (square of) the derivative and the second derivative of the
arbitrary function $W(x)$, usually referred to as the
superpotential.

Therefore, a {\underline {given}} classical Hamiltonian: \quad
$H={1\over 2}p^2+V(x)$,  admits an extension to a ${\cal N}=2$
supersymmetric partner $H_S$ if and only if the superpotential
satisfies
\begin{equation}
{1\over 2}\frac{dW}{dx}\cdot\frac{dW}{dx}=V(x)\label{eq:inc}
\end{equation}
Note that $\frac{d^2W}{dx^2}$ enters in $H_S$ as the expectation
value in Grassman states and disappears in a purely bosonic
setting.

\noindent Let us now consider the Hamiltonian for the \lq\lq
flipped" potential $V(x)=-U(x)$ and the associated Hamilton-Jacobi
equation:
\[
H_F={1\over 2}p^2+U(x) \quad ; \qquad \frac{\partial S}{\partial
t}+H_F(\frac{\partial S}{\partial x},x)=0 \qquad .
\]
The time-independence of the Hamiltonian suggests solutions of the
form $S(x,t)=-Et+W(x)$, leading to the reduced HJ equation:
\begin{equation}
{1\over 2}\frac{dW}{dx}\cdot\frac{d
W}{dx}+U(x)=E\label{eq:incr}\qquad .
\end{equation}
Therefore, the superpotential is no more than the Hamilton
characteristic function for $E=0$ of the mechanical system with
flipped potential. In sum, to find the superpotential, allowing
for the supersymmetric extension of a classical mechanical system,
one must solve a related Hamilton-Jacobi equation, see Reference
\cite{Gol}.

\noindent In general, for any $E$, the Hamilton characteristic
function is:
\begin{equation}
W(x;E)=\pm\int \, dx \sqrt{2(E-U(x))} \qquad \qquad .
\end{equation}
The energy $E$ trajectories satisfy the ODE
\begin{equation}
\frac{dx}{dt}=\pm \frac{dW}{dx}=\pm \sqrt{2(E-U(x))}\quad
\Rightarrow\quad \pm \int \, \frac{dx}{\sqrt{2(E-U(x))}}=t+t_0
\end{equation}

\subsection{One-dimensional ${\cal N}=2$ SUSY quantum mechanics}

Canonical quantization of the above system to obtain the analogous
${\cal N}=2$ quantum supersymmetric system proceeds as follows,
see, e.g., References \cite{Cas}, \cite{Boris}, \cite{Ioffe} and
\cite{Ioffe2}:

\noindent {\bf 1.} Replace Poisson brackets by commutators for the
bosonic variables and anticommutators for the fermionic variables:
\[
[\hat{x} , \hat{p}]=\hat{x}  \hat{p}-\hat{p} \hat{x} =
i\bar{\hbar} \qquad , \qquad \{\hat{\theta}_\alpha ,
\hat{\theta}_\beta \}=\hat{\theta}_\alpha \hat{\theta}_\beta +
\hat{\theta}_\beta \hat{\theta}_\alpha = -\bar{\hbar}
\delta_{\alpha\beta} \qquad ,
\]
where the non-dimensional Planck constant
${\bar{\hbar}}=\frac{\lambda\hbar}{m^{1\over 2}k^{3\over 2}}$ has
been introduced.

\noindent {\bf 2.} We choose the coordinate representation for the
bosonic variables but the classical Grassman variables become
Fermi operators in the quantum domain: $\hat{p} =
{{\bar{\hbar}}\over i} { d\over dx}$, $\hat{x}=x$,
$\hat{\theta}_1=\sqrt{{\bar{\hbar}}}\psi_1$,
$\hat{\theta}_2=\sqrt{{\bar{\hbar}}}\psi_2$, $ \{\psi_1,\psi_2\}=0
$.

\noindent The Fermi operators are represented on the Euclidean
spinors in ${\mathbb R}^2$ by the anti-Hermitian $2\times 2$ Pauli
matrices:
\[
\psi_1=\frac{i}{\sqrt{2}}\sigma_1 \ , \quad
\psi_2=-\frac{i}{\sqrt{2}}\sigma_2 \ , \quad
\hat{\theta}_1^2=\hat{\theta}^2_2=-\frac{\bar\hbar}{2}\left(\begin{array}{cc}
1 & 0 \\ 0 & 1 \end{array}\right)\ , \quad
\{\hat{\theta}_1,\hat{\theta}_2\}=0 \  ,\quad [
\hat{\theta}_1,\hat{\theta}_2]=i\bar\hbar\sigma_3\quad .
\]


\noindent {\bf 3.} The quantum supercharges,
$\hat{Q}_1=-\hat{\theta}_1\bar\hbar\frac{d}{dx}-i\hat{\theta}_2\frac{dW}{dx}$,
$\hat{Q}_2=\hat{\theta}_2\bar\hbar\frac{d}{dx}+i\hat{\theta}_1\frac{dW}{dx}$,
are
\[
\hat{Q}_1=-i\sqrt{\frac{\bar{\hbar}}{2}}\left(\begin{array}{cc} 0
& \bar{\hbar}\frac{d}{dx}-\frac{dW}{dx} \\
\bar{\hbar}\frac{d}{dx}+\frac{dW}{dx} & 0
\end{array}\right) \ , \qquad
\hat{Q}_2=\sqrt{\frac{\bar{\hbar}}{2}}\left(\begin{array}{cc} 0 &
-\bar{\hbar}\frac{d}{dx}+\frac{dW}{dx} \\
\bar{\hbar}\frac{d}{dx}+\frac{dW}{dx} & 0
\end{array}\right)
\]
and satisfy the quantum  algebra:
$\{\hat{Q}_1,\hat{Q}_1\}=\{\hat{Q}_2,\hat{Q}_2\}=2\bar{\hbar}\hat{H}_S
$, $\{\hat{Q}_1,\hat{Q}_2\}=0$,
$[\hat{Q}_1,\hat{H}_S]=[\hat{Q}_2,\hat{H}_S]=0$,
with the quantum SUSY Hamiltonian:
\[
\hat{H}_S=\left(\begin{array}{cc} \hat{h}^{(0)} & 0 \\ 0 &
\hat{h}^{(1)}\end{array}\right)={1\over 2}\left(\begin{array}{cc}
-\bar{\hbar}^2\frac{d^2}{dx^2}+\frac{dW}{dx}\cdot\frac{dW}{dx}-\bar{\hbar}\frac{d^2W}{dx^2}
& 0 \\ &  \\ 0 &
-\bar{\hbar}^2\frac{d^2}{dx^2}+\frac{dW}{dx}\cdot\frac{dW}{dx}+\bar{\hbar}\frac{d^2W}{dx^2}\end{array}\right)\qquad
.
\]
It is also interesting to work with non-hermitian supercharges
$\hat{Q}_\pm={1\over 2}(\hat{Q}_1 \pm i \hat{Q}_2)$,
\[
\hat{Q}_+=-i\sqrt{\frac{\bar{\hbar}}{2}}\left(\begin{array}{cc} 0
&
\bar{\hbar}\frac{d}{dx}-\frac{dW}{dx} \\
0 & 0
\end{array}\right) \ , \ \hat{Q}_-=-i\sqrt{\frac{\bar{\hbar}}{2}}\left(\begin{array}{cc} 0 &
0 \\
\bar{\hbar}\frac{d}{dx}+\frac{dW}{dx} & 0
\end{array}\right)\qquad ,
\]
and reshuffle the quantum superalgebra in the form: $\{\hat{Q}_+ ,
\hat{Q}_- \}=2 \bar{\hbar} \hat{H}_S$,
$[\hat{Q}_+,\hat{H}_S]=[\hat{Q}_-,\hat{H}_S]=0$.


\noindent {\bf 4.} The quantum Hamiltonian is a block-diagonal
$2\times 2$ matrix differential operator $\hat{h}^{(f=0)}$ and
$\hat{h}^{(f=1)}$ are ordinary Schr\"odinger operators acting
respectively on the subspaces of the Hilbert superspace labeled by
the eigenvalues of the Fermi number operator:
\[
\hat{f}=\hat{\theta}_-\cdot\hat{\theta}_+=\frac{\bar\hbar}{2}\cdot
\left(\begin{array}{cc} 0 & 0 \\ 0 & 1\end{array}\right) \qquad ,
\qquad \hat{\theta}_\pm={i\over 2}(\hat{\theta}_1\pm
i\hat{\theta}_2)
\]

\noindent {\bf 5.} Wave functions in the subspaces with zero and
one Fermi number annihilated respectively by $\hat{Q}_+$ and
$\hat{Q}_-$: $\hat{Q}_+\Psi_0^{(0)}(x)=0$,
$\hat{Q}_-\Psi_0^{(1)}(x)=0$,
are eigenfunctions of the Hamiltonian of zero energy. Therefore,
\[ \Psi_0^{(0)}(x)=C{\footnotesize \left(
\begin{array}{c} {\rm exp}[\bar{\hbar}^{-1} W(x)]
\\  0
\end{array} \right)} \qquad  , \qquad  \Psi_0^{(1)}(x)=C{\footnotesize \left(
\begin{array}{c} 0
\\  {\rm exp}[-\bar{\hbar}^{-1}W(x)]
\end{array} \right)}
\]
are the ground states of the supersymmetric quantum system if they
are normalizable: $ \int_{{\mathbb R}}\,dx\,
e^{2W(x)\bar{\hbar}^{-1}}<+\infty$ \ \ ${\underline{\rm or}}$
\\$\int_{{\Bbb R}}\,dx\, e^{-2W(x)\bar{\hbar}^{-1}}<+\infty$. Note that either $\Psi_0^{(0)}$ or $\Psi_0^{(1)}$ can be
normalizable.

\subsection{The two-fold way to supersymmetric quantum mechanics}

Given a physical system, the issue of building the associated
supersymmetric quantum mechanics can be addressed in two different
ways.

\noindent $\bullet$ {\it Quantization of a classical
supersymmetric system.} In the first method, it is assumed that
the classical supersymmetric extension has been performed. The
identification of the classical superpotential requires that we
must solve the ODE
\[
{1\over 2} {dW\over dx} {dW\over dx} = V(x) \qquad ,
\]
the time-independent Hamilton-Jacobi equation (\ref{eq:incr}) for
zero energy and flipped potential energy. This idea has been
applied to integrable but not separable systems with two degrees
of freedom in Reference \cite{Ioffe3}. Canonical quantization, as
in the previous Section, provides all the interactions in the
quantum system
\[
{\hat V}^{(0)}(x)={1\over 2}{dW\over dx}{dW\over
dx}+{\bar{\hbar}\over 2} {d^2W\over dx^2} \qquad , \qquad  {\hat
V}^{(1)}(x) = {1\over 2}{dW\over dx}{dW\over dx}-{\bar{\hbar}\over
2} {d^2W\over dx^2}
\]
in terms of the Hamilton characteristic function.

\noindent $\bullet$ {\it Supersymmetrization of a quantum system.}
The identification of the \lq\lq quantum" superpotential would
require one to solve one of the two Riccati differential equations
\begin{equation}
{1\over 2} {d\hat{W}\over dx}{d\hat{W}\over
dx}\pm{\bar{\hbar}\over 2}{d^2\hat{W}\over dx^2} = V(x)\qquad ,
\label{eq:ric1}
\end{equation}
the sign marking the subspace where the the potential energy $V$
is expected to act. There is no dependence on the Planck constant
in the potential energy of any physically significant mechanical
system. Therefore, we change the strategy and look for
superpotentials that solve the Poisson equation:
\begin{equation}
\pm{d^2\hat{W}_P\over dx^2}(x)=V(x) \qquad \qquad ,
\label{eq:poiss}
\end{equation}
with the same criterion for the signs. Physically, this means that
the Yukawa interactions provide the potential energy at stake.
Mathematically, the solution of the Poisson equation
(\ref{eq:poiss}) $\hat{W}_P$ provides a solution to a pair of
related Riccati equations (\ref{eq:ric2}):
\begin{equation}
{1\over 2} {d\hat{W}_P\over dx}{d\hat{W}_P\over
dx}+{\bar{\hbar}\over 2}{d^2\hat{W}_P\over dx^2} =
\hat{V}^{(0)}(x)\qquad , \qquad {1\over 2} {d\hat{W}_P\over
dx}{d\hat{W}_P\over dx}-{\bar{\hbar}\over 2}{d^2\hat{W}_P\over
dx^2} = \hat{V}^{(1)}(x)\qquad , \label{eq:ric2}
\end{equation}
for other related potential energies: $\hat{V}^{(0)}(x)$,
$\hat{V}^{(1)}(x)$. Once again, the datum is $V(x)$ in
(\ref{eq:poiss}) from which $\hat{V}^{(0)}(x)$, $\hat{V}^{(1)}(x)$
are derived.

\section{Examples: Anharmonic oscillators of sixth-order}
To put these ideas to work, we choose as examples one-dimensional
oscillators with terms proportional to $x^4$ {\underline and}
$x^6$ in the potential energy. Papers, reviews and even books
dealing with the $x^4$ case abound. We shall discuss the $x^6$
case because it provides a splendid arena to disentangle two
effects, instantons and spontaneous supersymmetry breaking, which
in the $x^4$ case come together. The potential energies are:
\begin{equation}
V(x;\lambda,k)=\frac{\lambda^2}{2k}x^2\left(x^2\pm\frac{k}{\lambda}\right)^2
\quad \qquad , \quad \qquad V(x)={1\over 2}\, x^2 (x^2\pm1)^2
\label{eq:6barr} \quad ,
\end{equation}
describing respectively a single (+ sign) or triple (- sign) well.
We shall only describe the first line of attack here from the
solution to the HJ equation (where the potential energy is not
found in the Yukawa interactions) and leave the Poisson route for
another publication.

\subsection{Quantization of classical supersymmetric sixth-order wells}

\subsubsection{Single well}

\noindent {\it 1.} {\it Supersymmetric classical mechanics.} The
solution to the HJ equation for $E=0$ and $U(x)= -{1\over 2}
x^2(x^2+1)^2 $ is:
\[
W(x)=\pm \int \, dx \, x(x^2+1)=\pm\left({x^4\over 4}+{x^2\over
2}\right) \qquad .
\]
The supersymmetric classical Hamiltonian and the supercharges
read:
\[
H_S={1\over 2} p^2+{1\over 2}x^2(x^2+1)^2\mp
i((3x^2+1)\theta_2\theta_1 \qquad , \qquad
Q_\alpha=p\theta_\alpha\mp
x(x^2+1)\varepsilon_{\alpha\beta}\theta_\beta \qquad .
\]
In the \lq\lq soul" of the related supersymmetric system with
flipped potential, the Hamilton characteristic function and the
trajectories are given analytically by hyperelliptic integrals:
\[
W(x;E)=\int \, dx \, \sqrt{x^6+2x^4+x^2+2E} \qquad , \qquad \int
\, \frac{dx}{\sqrt{x^6+2x^4+x^2+2E}}=t+t_0 \qquad .
\]
For $E=0$, there is only one constant trajectory, where the
particle sits on the top of the potential: $\bar{x}(t)=0$, which
is also the unique BPS trajectory of the supersymmetric classical
system.

\noindent {\it 2.} {\it Supersymmetric quantum mechanics.} The
quantum supercharges are: {\footnotesize\begin{equation}
\hat{Q}_1=-i\sqrt{\frac{\bar{\hbar}}{2}}\left(\begin{array}{cc} 0
& \bar{\hbar}\frac{d}{dx}\mp x(x^2+1) \\
\bar{\hbar}\frac{d}{dx}\pm x(x^2+1) & 0
\end{array}\right) \hspace{0.2cm} ,\hspace{0.2cm}
\hat{Q}_2=\sqrt{\frac{\bar{\hbar}}{2}}\left(\begin{array}{cc} 0 &
-\bar{\hbar}\frac{d}{dx}\pm x(x^2+1) \\
\bar{\hbar}\frac{d}{dx}\pm x(x^2+1) & 0
\end{array}\right)\label{eq:qsup0}\hspace{0.2cm} ,
\end{equation}}
and the potential energies arising in $H_S$ read:
\begin{equation}
\hat{V}^{(0)}(x) = {1\over 2} (x^2(x^2+1)^2 \pm
\bar{\hbar}(3x^2+1))\qquad , \qquad \hat{V}^{(1)}(x) = {1\over 2}
(x^2(x^2+1)^2 \mp \bar{\hbar}(3x^2+1)) \qquad . \label{eq:qpot0}
\end{equation}
Thus, the zero energy ground states are:
\[
\Psi_0^{(0)}(x)=C\left(\begin{array}{c} {\rm exp}\{\pm\frac{({x^4
\over 4}+{x^2\over 2})}{\bar{\hbar}}\}
 \\ 0
\end{array}\right) \qquad , \qquad \Psi_0^{(1)}(x)=C\left(\begin{array}{c} 0 \\
{\rm exp}\{\frac{\mp({x^4\over 4}+{x^2\over 2})}{\bar{\hbar}}\}
\end{array}\right) \qquad .
\]
The supersymmetric quantum system has always one ground (BPS)
state and supersymmetry is unbroken: if we choose $W={x^4\over
4}+{x^2\over 2}$ as the superpotential, the ground state belongs
to the Fermi subspace - $\Psi_0^{(0)}$ is not normalizable-, the
choice of $W={x^2\over 2}+{x^4\over 4}$ forces a Bosonic ground
state whereas $\Psi_0^{(1)}$ becomes non-normalizable.

One can guess the energy and the type of eigen-function of the
next energetic states by looking at the \lq\lq effective"
potentials:
\[
V_+(x) = {1\over 2} (x^2(x^2+1)^2 - \bar{\hbar}(3x^2+1))\qquad ,
\qquad V_-(x) = {1\over 2} (x^2(x^2+1)^2 + \bar{\hbar}(3x^2+1))
\qquad ,
\]
either $\hat{V}^{(0)}$ or $\hat{V}^{(1)}$ depending on the choice
of $W$. The critical points of $V_\pm(x)$ are: $x_0=0 $, $x_1^\pm
= \pm\sqrt{-\frac{2 - \sqrt{1\pm 9\bar{\hbar}}}{3}}$, $x_2^\pm =
\pm\sqrt{\frac{-2 + \sqrt{1\pm 9\bar{\hbar}}}{3}}$,
\[
{d^2V_\pm\over dx^2}(x_0)=1\mp 3\bar{\hbar} \quad , \quad
{d^2V_\pm\over dx^2}(x_1^\pm)=4(1\pm 3\bar{\hbar}+\sqrt{1\pm
9\bar{\hbar}})\quad , \quad {d^2V_\pm\over dx^2}(x_2^\pm)=4(1\pm
3\bar{\hbar}-\sqrt{1\pm 9\bar{\hbar}}) \qquad .
\]
$x_0$ is a minimum of $V_+$ if $\bar{\hbar}<{1\over 3}$ and
becomes a maximum otherwise. $x_1^\pm$ are always imaginary roots
but $x_2^\pm$ are real and become minima of $V_+$ for
$\bar{\hbar}>{1\over 3}$.

\noindent\begin{figure}[htbp] \centerline{
\epsfig{file=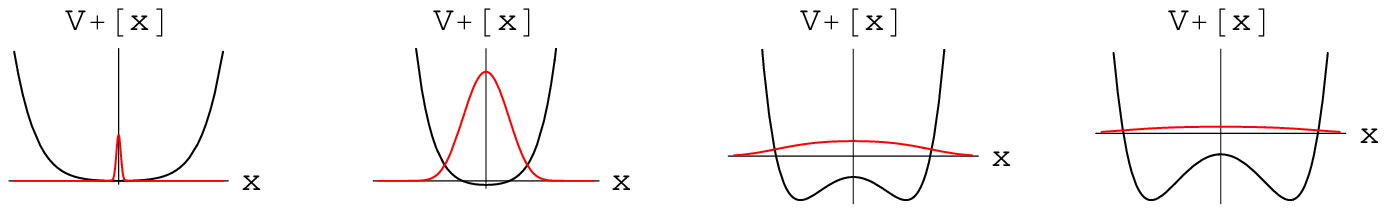,height=1.8cm}} \caption{\small Potential
energy ${\hat V}_+$ and BPS ground state $\Psi_0^G(x)$ (red) for:}
\centerline{\small (a) ${\bar \hbar}=0.001$, (b) ${\bar
\hbar}=0.1$, (c) ${\bar \hbar}= 2$, (d) ${\bar \hbar}=4$. }
\end{figure}

There is a unique minimum for $V_-$, $x_0$, and the wave function
of the first level over the ground state is well approximated by a
Gaussian around it:
\begin{equation}
\Psi_-^{E_1^-}(x)\simeq\left(\frac{\omega_-}{\bar{\hbar}\pi}\right)^{{1\over
4}}\left(\begin{array}{c}0 \\ {\rm exp}\{-{\omega_-\over 2
\bar{\hbar}}x^2\}\end{array}\right) \qquad , \qquad
\omega_-=\sqrt{1+3\bar{\hbar}} \qquad , \qquad E_-^{E_1^-}\simeq
{\bar{\hbar}\over 2}(1+\omega_-)\qquad .
\end{equation}
The supersymmetric partner state in the subspace of $\Psi_0^G$ is
obtained by acting on $\Psi_-^{E_1^-}$ with $\hat{Q}_+$:
\begin{equation}
\Psi_+^{E_1^+}(x)=\hat{Q}_+\Psi_-^F(x)=\left(\frac{\omega_-\bar{\hbar}}{4\pi}\right)^{{1\over
4}}\left(\begin{array}{c}(x^3+(1+\omega_-)x){\rm
exp}[-{\omega_-\over\bar{\hbar}}x^2] \\ 0 \end{array}\right) \ , \
E_1^+=E_1^- \qquad .
\end{equation}

\noindent\begin{figure}[htbp] \centerline{
 \epsfig{file=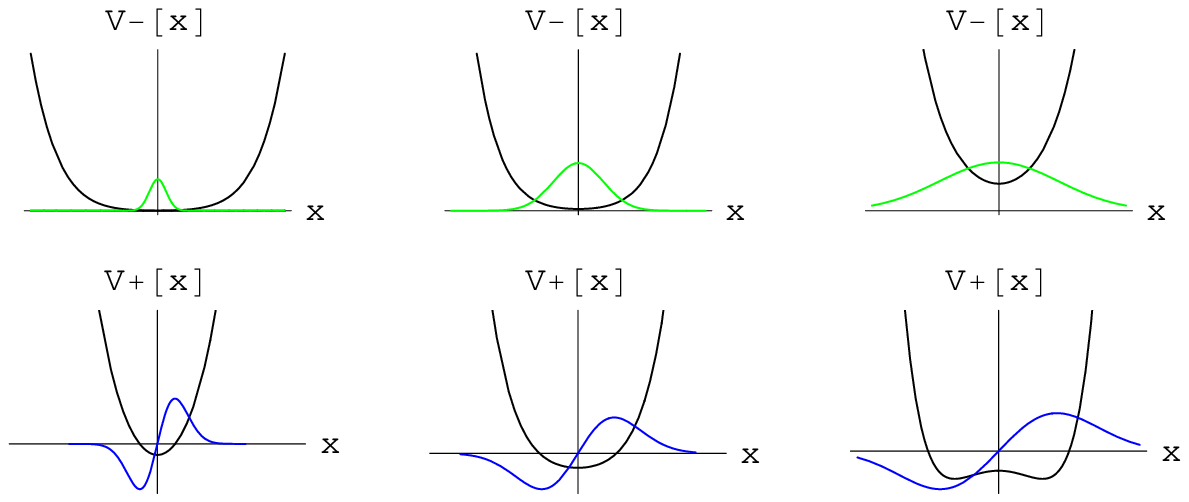,height=4cm}}
 \caption{ \small Potential energies ${\hat V}_{\pm}$. Degenerate in energy $\Psi_-^{E_1^-}(x)$ (green)
 }

 \centerline {\small and $\Psi_+^{E_1^+}(x)$ (blue) wave functions: (a) ${\bar \hbar}=0.01$, (b) ${\bar \hbar}=0.1$, (c) ${\bar \hbar}=
1$.}
\end{figure}

\noindent {\it 3.} {\it Zero-energy ground state.} The dependence
of $\Psi_0^G(x)={\rm exp}[-{({x^4\over 4}+{x^2\over
2})\over\bar{\hbar}}]$ on $\bar{\hbar}$ is rather involved and can
be described analytically through the asymptotic behavior when
$\bar{\hbar}\rightarrow\bar{\hbar}_c$ and $\bar{\hbar}_c=0$ is the
classical value:
\begin{eqnarray*}
&&{\rm exp}[-\frac{1}{2\bar{\hbar}}({x^4\over
2}+x^2)]\simeq\lim_{\bar{\hbar}_c\rightarrow 0}{\rm
exp}[-\frac{1}{\bar{\hbar}_c}({x^4\over
2}+x^2)]\left\{1+{1\over\bar{\hbar}_c^2}\cdot\left(\frac{2\bar{\hbar}}{({x^4\over
2}+x^2)}-\bar{\hbar}_c\right)\right.\\&+& \left.
\frac{1-2\bar{\hbar}_c}{2\bar{\hbar}_c^4}\cdot\left(\frac{2\bar{\hbar}}{({x^4\over
2}+x^2)}-\bar{\hbar}_c\right)^2
+\frac{1+6\bar{\hbar}_c(\bar{\hbar}_c-1)}{6\bar{\hbar}_c^6}\cdot\left(\frac{2\bar{\hbar}}{({x^4\over
2}+x^2)}-\bar{\hbar}_c\right)^3 +\cdots\right\} \qquad .
\end{eqnarray*}

\noindent It is also interesting to analyze how the norm of the
BPS ground state $\Psi_0^G(x)$ depends on $\bar{\hbar}$:
\begin{equation}
N(\bar{\hbar})= \int \, dx \, {\rm
exp}\left\{-\frac{x^4}{2\bar{\hbar}}-\frac{x^2}{\bar{\hbar}}\right\}
= \sqrt{\bar{\hbar}}\int \, dz \, {\rm
exp}\left\{-\frac{\bar{\hbar}z^4}{2}-z^2\right\} \qquad , \qquad
x=\sqrt{\bar{\hbar}}z \qquad .
\end{equation}
This non-gaussian integral is no more than the partition function
${\cal Z}(\bar{\hbar})=N(\bar{\hbar})$ of a QFT system in
(0+0)-spacetime dimensions and Lagrangian \cite{Bor}:
\begin{equation}
L= -{1\over 2} \varphi^2 -{\lambda\over 4!} \varphi^4 \qquad ,
\qquad z={\varphi \over \sqrt{2}}\quad , \quad
\lambda=3\bar{\hbar} \qquad .
\end{equation}
The partition function can be expressed as a series in
$\bar{\hbar}$,
\begin{equation}
{\cal Z}(\bar{\hbar})= \sum_{m=0}^{\infty}
\int_{-\infty}^{\infty}\, d z \, {(-3 \bar{\hbar})^m \over (4!)^m
m!}\,z^{4m}\,  e^{-z^2} \qquad , \qquad \int_{-\infty}^{\infty} \,
dz \, z^{4m} \, e^{-z^2} = \frac{(4m)!}{(2m)!2^{2m}} \cdot
\sqrt{\pi} \quad ,
\end{equation}
by performing infinite Gaussian integrals:
\begin{equation}
{{\cal Z}(\bar{\hbar})\over \sqrt{\bar{\hbar} \pi} } =
\sum_{m=0}^{\infty} {(-3 \bar{\hbar})^m \over (4!)^m m!} \cdot {
(4 m)! \over (2 m)! 2^{2m}} = 1 -{1\over 8} (3 \bar{\hbar}) + {5
\cdot 7 \over 3 \cdot 2^7} (3 \bar{\hbar})^2 - {5 \cdot 7 \cdot 11
\over 3 \cdot 2^{10}} (3 \bar{\hbar})^3 + \cdots  \qquad .
\label{eq:ass}
\end{equation}
The expansion (\ref{eq:ass}) of the partition function shows an
essential singularity at $\bar{\hbar}=0$ -the classical limit- and
it is an asymptotic series. The best approximation to the integral
is reached by keeping a number of terms $m_0$ such that the
quotient between two consecutive terms is of the order of one:
\[
\left|{a_{m_0+1} \over a_{m_0}} \right| = { (4 m_0 +3) (4 m_0 +1)
\over 4! (m_0+1) } |3 \bar{\hbar}| \approx \bar{\hbar} 2 m_0
\approx 1 \Rightarrow m_0 \approx {1\over 2 \bar{\hbar}} \qquad ,
\]
and the error assumed by neglecting higher-order terms is bounded
by ${\rm exp}[-{1\over 2 \bar{\hbar}}]$.

It is tempting to explain the pictorial description of the series
using Feynman diagram technology. Writing the partition function
in the form,
\begin{equation}
{{\cal Z}[\lambda]\over \sqrt{ \lambda/3}} = \sum \int {
\overbrace{ {(-\lambda) \varphi^4 \over 4!} \times \cdots \times
{(-\lambda) \varphi^4 \over 4!} } \over m!} e^{ \left( -{1\over 2}
\varphi^2 \right)} d \varphi \quad , \label{eq:feyn}
\end{equation}
one discovers the following Feynman rule: there is a single
tetravalent vertex with a factor $(-\lambda)$. The lower-order
terms in the series (\ref{eq:ass}) correspond to the weights of
the vacuum diagrams - the $-\lambda$ factor of the vertex divided
by the combinatorial factor, the number of equivalent graphs of
the same topological type- up to second order in perturbation
theory shown in the next Table.

\begin{table}[h]
\begin{center}
\begin{tabular}{cccccc} \\ \hline
\textit{ Vacuum graph} &  & \textit{ Weight }&\textit{ Vacuum graph} &  & \textit{ Weight }\\
\hline \\ \parbox{2.5cm}{\epsfig{file=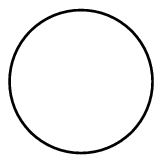,width=0.5cm}} &
$\rightarrow$ & { $1$} & & &  \\ & & \\

\parbox{2.5cm}{\epsfig{file=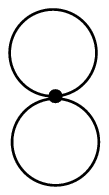,width=0.5cm}}
& $\rightarrow$ &  $ - {\displaystyle{ \lambda \over 2^3} }$ &
\parbox{2.5cm}{\epsfig{file=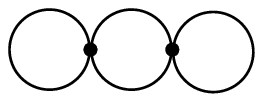,width=1.5cm}}
& $\rightarrow$ &  ${\displaystyle{ \lambda^2 \over 2^4}}$\\& & \\
\parbox{2.5cm}{\epsfig{file=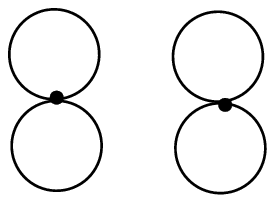,width=1.2cm}}
& $\rightarrow$ &  $ {\displaystyle{ \lambda^2 \over 2^7 }} $ &
\parbox{2.5cm}{\epsfig{file=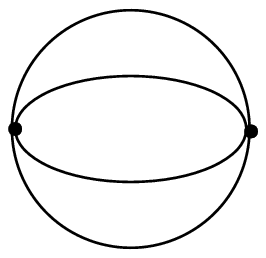,width=0.9cm}}
& $\rightarrow$ &  $ {\displaystyle{ \lambda^2 \over 3 \cdot 2^4}} $ \\
[0.5cm] \hline
\end{tabular}
\end{center}
\end{table}

\subsubsection{Triple well}

\noindent {\it 1.}  {\it Supersymmetric classical mechanics.} The
solution to the HJ equation for $E=0$ and $U(x)=-{1\over
2}x^2(x^2-1)^2$ is:
\[
W(x)=\pm\int \, dx \,
x(x^2-1)=\pm\left(\frac{x^4}{4}-\frac{x^2}{2}\right) \qquad .
\]
The superpotential is thus the \lq\lq sombrero" potential. The
supersymmetric classical Hamiltonian and the supercharges read:
\[
H_S={1\over 2} p^2+{1\over 2}x^2(x^2-1)^2\mp
i((3x^2-1)\theta_2\theta_1 \qquad , \qquad
Q_\alpha=p\theta_\alpha\mp x(x^2-1)\varepsilon_{\alpha\beta}
\theta_\beta \qquad .
\]
Although feasible, we shall not attempt to search for trajectories
with non-null Grassman degrees of freedom.

It is interesting, instead, to look at solutions in the \lq\lq
body" of the related supersymmetric system with flipped potential
because of their r$\hat{\rm o}$le in the quantum $H_S$ system. The
Hamilton characteristic function and the trajectories are given
analytically by hyperelliptic integrals:
\[
W(x;E)=\int \, dx \, \sqrt{x^6-2x^4+x^2+2E} \qquad , \qquad \int
\, \frac{dx}{\sqrt{x^6-2x^4+x^2+2E}}=t+t_0 \qquad .
\]
For $E=0$, the integrations are easily performed and two kinds of
trajectories are found:

\noindent $\bullet$ Constant trajectories, where the particle sits
on the top of the potential: \quad $\bar{x}(t)=0$, $\bar{x}(t)=\pm
1$.

\noindent $\bullet$ Trajectories where the particle starts from a
maximum of the potential at $t=-\infty$ and slowly moves to reach
$x=\pm\infty$ (infinite action) or another maximum (finite action)
at $t=\infty$.
\[
x^2>1: \ \bar{x}^2(t)={1\over 1-e^{\pm 2(t+t_0)}} \hspace{0.5cm} ,
\hspace{0.5cm} x^2<1, \,\, {\underline{\rm instanton}}: \
\bar{x}(t)=\pm \frac{1}{\sqrt{1+e^{\pm 2(t+t_0)}}} \qquad .
\]

\noindent\begin{figure}[htbp] \centerline{
 \epsfig{file=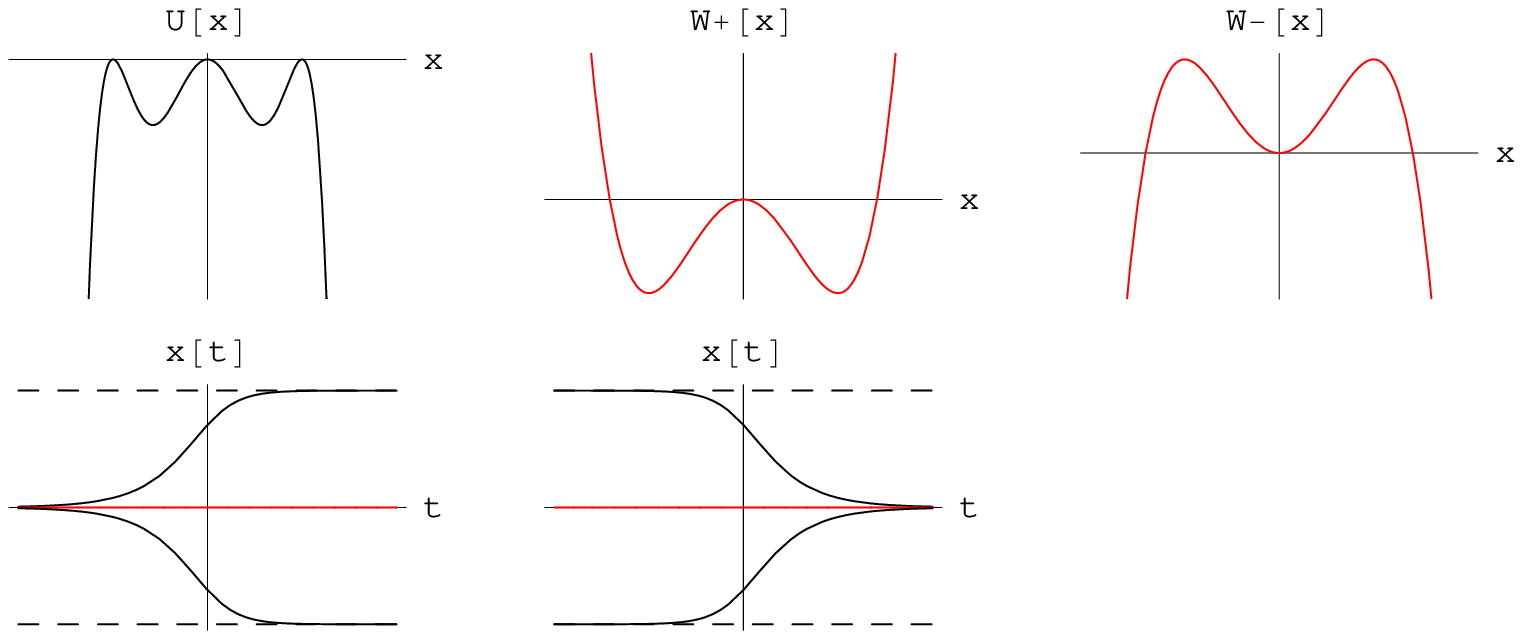,height=3.5cm}}
 \caption{ \small (a) Potential energy $U(x)$. (b) Hamilton characteristic function (superpotential)
 $W^{\pm}(x)$.}
\centerline{\small (c) Zero-energy, finite action, trajectories
(instantons). }
\end{figure}
The constant trajectories are special due to the fact that they
are also zero energy (BPS) classical solutions to $H_S$ because
the classical supercharges $Q_\pm=(p\mp ix(x^2-1))\theta_\pm$ are
annihilated by them for any value of $\theta_\pm$.

\noindent {\it 2.} {\it Supersymmetric quantum mechanics.} The
quantum supercharges are: {\footnotesize\begin{equation}
\hat{Q}_1=-i\sqrt{\frac{\bar{\hbar}}{2}}\left(\begin{array}{cc} 0
& \bar{\hbar}\frac{d}{dx}\mp x(x^2-1) \\
\bar{\hbar}\frac{d}{dx}\pm x(x^2-1) & 0
\end{array}\right) \hspace{0.2cm} ,\hspace{0.2cm}
\hat{Q}_2=\sqrt{\frac{\bar{\hbar}}{2}}\left(\begin{array}{cc} 0 &
-\bar{\hbar}\frac{d}{dx}\pm x(x^2-1) \\
\bar{\hbar}\frac{d}{dx}\pm x(x^2-1) & 0
\end{array}\right)\label{eq:qsup}\hspace{0.2cm} ,
\end{equation}}
and the potential energies arising in $H_S$ read:
\begin{equation}
\hat{V}^{(0)}(x) = {1\over 2} (x^2(x^2-1)^2 \pm
\bar{\hbar}(3x^2-1))\qquad , \qquad \hat{V}^{(1)}(x) = {1\over 2}
(x^2(x^2-1)^2 \mp \bar{\hbar}(3x^2-1)) \qquad . \label{eq:qpot}
\end{equation}
Thus, the zero-energy ground states are:
\[
\Psi_0^{(0)}(x)=C\left(\begin{array}{c} {\rm exp}\{\pm\frac{({x^4
\over 4}-{x^2\over 2})}{\bar{\hbar}}\}
 \\ 0
\end{array}\right) \qquad , \qquad \Psi_0^{(1)}(x)=C\left(\begin{array}{c} 0 \\
{\rm exp}\{\frac{\mp({x^4\over 4}-{x^2\over 2})}{\bar{\hbar}}\}
\end{array}\right) \qquad .
\]
The supersymmetric quantum system always has one ground (BPS)
state and supersymmetry is unbroken: if we choose $W={x^4\over
4}-{x^2\over 2}$ as superpotential the ground state belongs to the
Fermi subspace - $\Psi_0^{(0)}$ is not normalizable-, the choice
of $W={x^2\over 2}-{x^4\over 4}$ forces a bosonic ground state
whereas $\Psi_0^{(1)}$ becomes non normalizable.

Nevertheless, despite unbroken supersymmetry this system has
instantons. To analyze the coexistence of these two phenomena one
needs to study how
\[
V_+(x) = {1\over 2} (x^2(x^2-1)^2 - \bar{\hbar}(3x^2-1))\qquad ,
\qquad V_-(x) = {1\over 2} (x^2(x^2-1)^2 + \bar{\hbar}(3x^2-1))
\]
evolve in response to changes in $\bar{\hbar}$. Note that either
$V_+$ or $V_-$ are either $\hat{V}^{(0)}$ or $\hat{V}^{(1)}$,
depending on the choice of $W$. The critical points of $V_\pm(x)$
are: $x_0=0$, $x_1^\pm = \pm\sqrt{\frac{2 - \sqrt{1\pm
9\bar{\hbar}}}{3}}$, $x_2^\pm = \pm\sqrt{\frac{2 + \sqrt{1\pm
9\bar{\hbar}}}{3}}$,
\[
{d^2V_\pm\over dx^2}(x_0)=1\mp 3\bar{\hbar} \quad , \quad
{d^2V_\pm\over dx^2}(x_1^\pm)=4(1\pm 3\bar{\hbar}-\sqrt{1\pm
9\bar{\hbar}})\quad , \quad {d^2V_\pm\over dx^2}(x_2^\pm)=4(1\pm
3\bar{\hbar}+\sqrt{1\pm 9\bar{\hbar}}) \qquad .
\]
\noindent\begin{figure}[htbp] \centerline{
 \epsfig{file=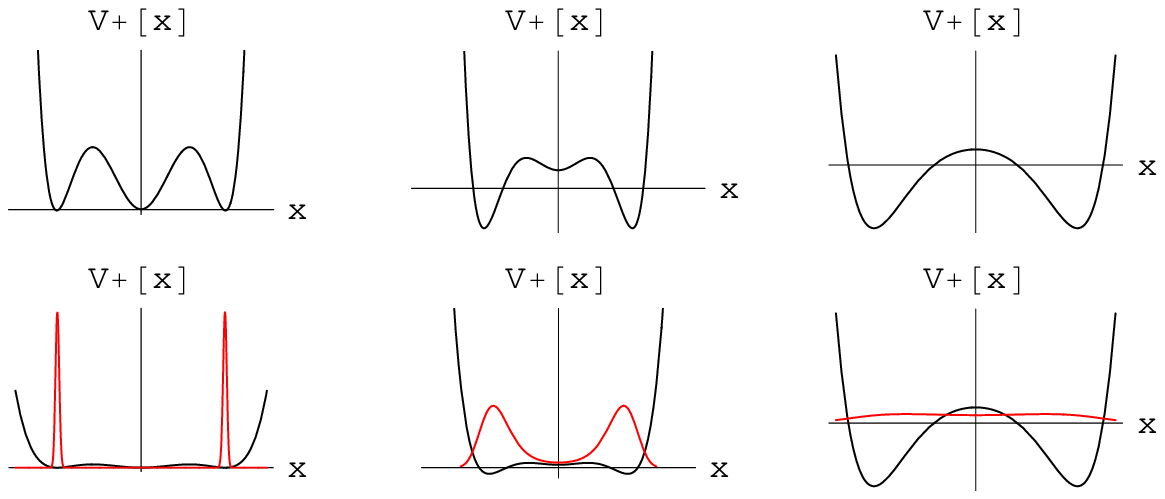,height=4cm}}
 \caption{ \small Potential energy $V_+$ and BPS wave function
 $\Psi_0^G(x)$ (red) plotted as functions of x }
 \centerline{ \small for several values of
 $\bar{\hbar}$:  (a) ${\bar \hbar}=0.001$, (b) ${\bar \hbar}=0.1$, (c) ${\bar \hbar}=
2$.}
\end{figure}

\noindent $x_2^\pm$ are always minima of $V_+(x)$, $x_0$ is a
minimum of $V_+$ if $\bar{\hbar}<{1\over 3}$ but becomes a maximum
if $\bar{\hbar}>{1\over 3}$, and $x_1^\pm$ are maxima of $V_+$ for
$\bar{\hbar}<{1\over 3}$, not anymore critical point for
$\bar{\hbar}>{1\over 3}$, see Figure 2. Therefore, because
$V_+(x_0)={\bar{\hbar}\over
2}>V_+(x_2^\pm)=\frac{2+\sqrt{1+9\bar{\hbar}}}{3}\left[{1-\sqrt{1+9\bar{\hbar}}\over
9}-2\bar{\hbar}\right]+{\bar{\hbar}\over 2}$, $x_0$ is a false
vacuum that decays to the true vacua $x_2^\pm$ when
$\bar{\hbar}<{1\over 3}$. The decay amplitude can be computed from
the classical bounce for the flipped potential, starting and
ending at $x_0$,  which is very well approximated by an
instanton-anti-instanton configuration for small values of
$\bar{\hbar}$. It is remarkable how well this behavior is
described by the ground state wave function $\Psi^G_0(x)$; even
more remarkable, $\Psi_0^G(x)$ also matches the expected behavior
for $\bar{\hbar}>{1\over 3}$ where there is no tunnel effect at
all, see again Figure 2.

$x_0$, however, is the absolute minimum of $V_-(x)$; if
$\bar{\hbar}<{1\over 9}$, $x_2^\pm$ are also minima of $V_-(x)$,
but $V_-(x_0)=-{\bar{\hbar}\over
2}<V_-(x_2^\pm)=\frac{2+\sqrt{1-9\bar{\hbar}}}{3}\left[{1-\sqrt{1-9\bar{\hbar}}\over
9}+2\bar{\hbar}\right]-{\bar{\hbar}\over 2}$. If
$\bar{\hbar}>{1\over 9}$ $x_0$ is the single critical point
(minimum) of $V_-(x)$. Therefore, the eigenfunction of the lowest
eigenvalue of the Schrodinger operator with potential energy
$V_-(x)$ is approximately a Gaussian centered at $x_0=0$:
\begin{equation}
\Psi_-^{E_1^-}(x)\simeq
\left(\frac{\omega_-}{\bar{\hbar}\pi}\right)^{{1\over 4}}\cdot
{\rm exp}\{-{\omega_-\over 2 \bar{\hbar}}x^2\}\ , \qquad
\omega_-=\sqrt{1+3\bar{\hbar}}\qquad , \qquad E_1^-\simeq
{\bar{\hbar}\over 2}(\omega_- -1) \qquad .\label{eq:nwf}
\end{equation}

\noindent $\Psi_-^{E_1^-}(x)$, the first eigenfunction of $H_s$
outside the kernel, lives in the subspace orthogonal to the
subspace of $\Psi^G(x)$. For $\bar{\hbar}<{1\over 9}$,
$\Psi_-^{E_1^-}$ grows from the decay of the false vacua $x_2^\pm$
ruled by instantons/anti-instantons now starting and ending at
$x_2^\pm$. Mathematica drawings of these wave functions are
offered in Figure 3. \noindent\begin{figure}[htbp] \centerline{
 \epsfig{file=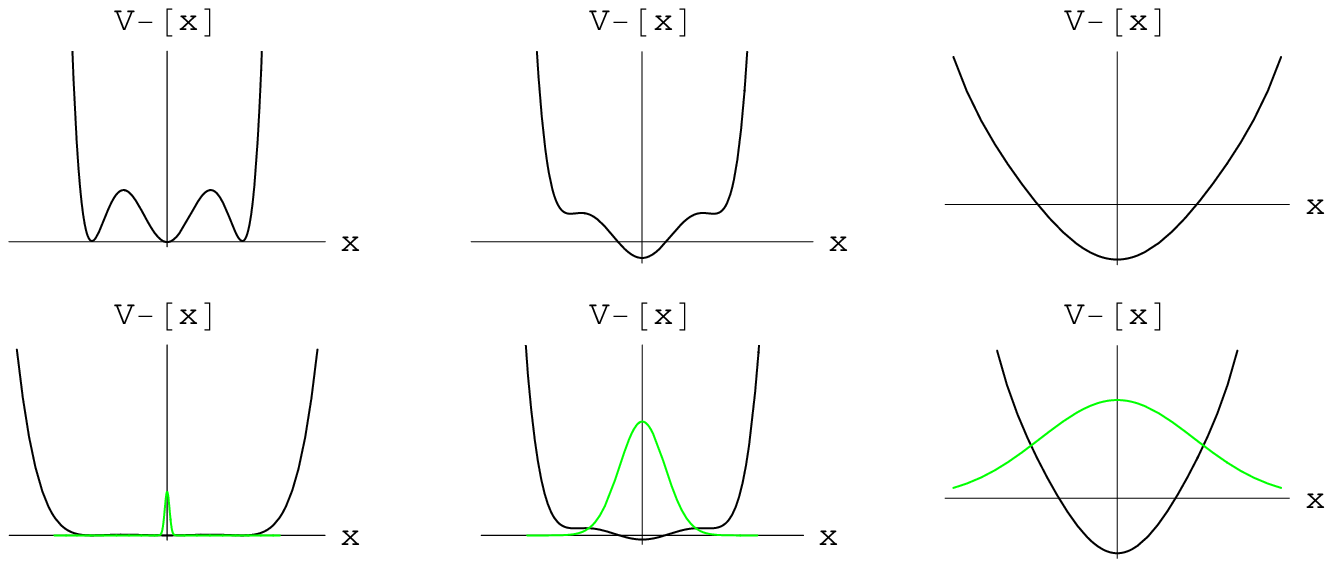,height=4cm}}
 \caption{ \small Potential energy  $ V_-(x)$ and wave function $\Psi_-^{E_1^-}(x)$ (green) plotted
 as functions of $x$ }
 \centerline{\small for several values of $\bar{\hbar}$:
(a) ${\bar \hbar}=0.001$, (b) ${\bar \hbar}=0.1$, (c) ${\bar
\hbar}= 1$. }
\end{figure}

Acting on $\Psi_-^{E_1^-}(x)$ with the supercharge operator
$\hat{Q}_+$, an approximate eigenfunction of $H_S$ is obtained in
the subspace of $\Psi^G(x)$. The supersymmetric partner of
$\Psi_-^{E_1^-}(x)$ is thus,
\begin{equation}
\Psi_+^{E_1^+}(x)=\hat{Q}_+\Psi_-^{E_1^-}(x)=\left(\frac{\omega_-\bar{\hbar}}{4\pi}\right)^{{1\over
4}}\cdot(x^3-(1-\omega_-)x){\rm exp}[-{\omega_-\over 2
\bar{\hbar}}x^2] \qquad \qquad , \qquad E_1^+=E_1^- \qquad ,
\end{equation}
and $E_1^+$ is the lowest-lying eigenvalue in the subspace of the
zero mode (ground state). Plots of these \lq\lq odd" wave
functions are shown in Figure 4  for several values of
$\bar{\hbar}$. The wave function has a node at the origin.

\noindent\begin{figure}[htbp] \centerline{
 \epsfig{file=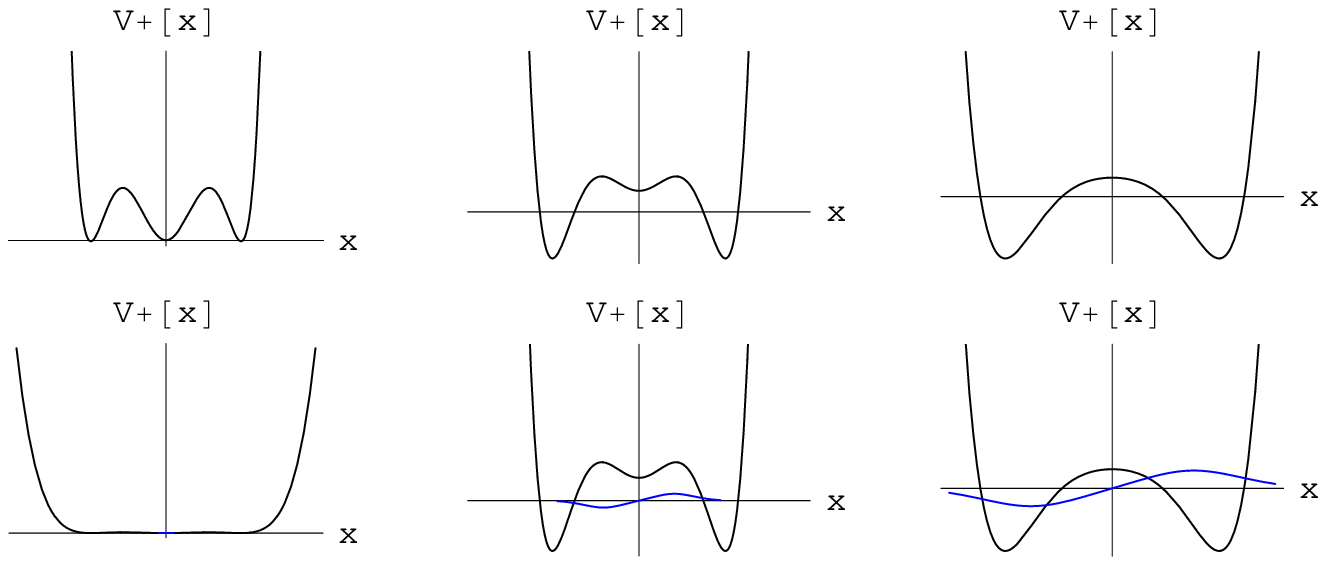,height=4cm}}
 \caption{ Potential energy $ V_+(x)$ and wave function $\Psi_+^{E_1^+}(x)$ (blue)
 as a function of $x$}
 \centerline{\small for several values of $\bar{\hbar}$:
(a) ${\bar \hbar}=0.001$, (b) ${\bar \hbar}=0.1$, (c) ${\bar
\hbar}= 1$. }
\end{figure}

\noindent {\it 3.} {\it Zero-energy ground state.} The dependence
of $\Psi_0^G(x)={\rm exp}[-{({x^4\over 4}-{x^2\over
2})\over\bar{\hbar}}]$ on $\bar{\hbar}$ is somewhat involved and
can be described analytically through the asymptotic behavior when
$\bar{\hbar}\rightarrow\bar{\hbar}_c$ and $\bar{\hbar}_c=0$ is the
classical value:
\begin{eqnarray*}
&&{\rm exp}[-\frac{1}{2\bar{\hbar}}({x^4\over
2}-x^2)]\simeq\lim_{\bar{\hbar}_c\rightarrow 0}{\rm
exp}[-\frac{1}{\bar{\hbar}_c}({x^4\over
2}-x^2)]\left\{1+{1\over\bar{\hbar}_c^2}\cdot\left(\frac{2\bar{\hbar}}{({x^4\over
2}-x^2)}-\bar{\hbar}_c\right)\right.\\&+& \left.
\frac{1-2\bar{\hbar}_c}{2\bar{\hbar}_c^4}\cdot\left(\frac{2\bar{\hbar}}{({x^4\over
2}-x^2)}-\bar{\hbar}_c\right)^2
+\frac{1+6\bar{\hbar}_c(\bar{\hbar}_c-1)}{6\bar{\hbar}_c^6}\cdot\left(\frac{2\bar{\hbar}}{({x^4\over
2}-x^2)}-\bar{\hbar}_c\right)^3 +\cdots\right\} \qquad .
\end{eqnarray*}

\noindent The norm of the BPS ground state $\Psi_0^G(x)$ is again
a non-Gaussian integral. Denoting $z ={x-1\over
\sqrt{\bar{\hbar}}}$, $2 z = \varphi$ and $ {3\over 4} \bar{\hbar}
= \beta $, we obtain:
\begin{equation}
N(\bar{\hbar}) = \int \, dx \, {\rm
exp}\left\{-\frac{x^4}{2\bar{\hbar}}+\frac{x^2}{\bar{\hbar}}\right\}
= e^{\bar{\hbar} \over 2} \sqrt{\bar{\hbar}}\int \, dz \, {\rm
exp}\left\{-\frac{\bar{\hbar}z^4}{2}- 2 \sqrt{\bar{\hbar}} z^3 - 2
z^2\right\}  \qquad . \label{eq:ngsw}
\end{equation}
$N(\bar{\hbar})={\cal Z}(\bar{\hbar})$ is the partition function
for the Euclidean $\lambda(\varphi)^4_0$-model with spontaneous
$x\rightarrow -x$ symmetry breaking in (0+0)-space time dimensions
and the Lagrangian:
\begin{equation}
L = -{1\over 2} \varphi^2 -{\beta \over 4!} \varphi^4 - {\sqrt{3
\beta} \over 3!} \varphi^3 \qquad .
\end{equation}
Performing infinite Gaussian integrals
\[
{{\cal Z}[\beta] \over e^{2 \beta \over 3} \sqrt{4 \beta/ 3} } =
\sum_{m=0}^{\infty} \sum_{2 k=0}^{\infty} \int { (-\beta)^m \over
(4!)^m m!} \cdot { (-\sqrt{3 \beta})^{2k} \over (3!)^{2k} (2k)!} \
\ \varphi^{4m+6k}\ \  e^{-{1\over 2} \varphi^2 } \ d \varphi
\]
one obtains the asymptotic $\bar{\hbar}$-expansion:
\begin{eqnarray}
&& {{\cal Z}[\beta] \over  e^{2 \beta \over 3} \sqrt{2 \beta
\pi/3}} = \sum_{m=0}^{\infty} \sum_{ k=0}^{\infty} { (-\beta)^m
\over (4!)^m m!} \cdot { (-\sqrt{3 \beta})^{2k} \over (3!)^{2k}
(2k)!} \cdot { (4m+6k)! \over (2m+3k)! 2^{2m+3k}}\label{eq:ass1}
\\ && \nonumber \\
&&=  1 \underbrace{-{\beta \over 8} +{ 5\cdot 7 \over 2^7 \cdot 3}
\beta^2 - {5 \cdot 7 \cdot 11 \over 2^{10} \cdot 3} \beta^3 +
\cdots}\underbrace{ - {3\cdot 5 \cdot 7 \over 2^6} \beta^2} +
\cdots  +
\underbrace{{5 \over 2^3} \beta + {5 \cdot 7 \cdot 11 \over 2^7}
\beta^2 + {5\cdot 7\cdot 11 \cdot 13 \cdot 17 \over 2^{10} \cdot
3} \beta^3 + \cdots}\nonumber
\end{eqnarray}
Again, the optimum value of the number of terms of $k$ type can be
estimated. Keeping a fixed but finite value of $m=m_0$ such that
$m_0<<k_0$, the quotient between two consecutive $k=k_0$ and
$k=k_0+1$ terms must be of the order of one:
\begin{eqnarray*}
\left|{a_{m_0+k_0+1} \over a_{m_0+k_0}} \right| &=&
\frac{1}{(2k_0+2)(2k_0+1)}\cdot\left[\frac{(4m_0+6k_0+6)(4m_0+6k_0+5)
\cdots(4m_0+6k_0+1)}{(2m_0+3k_0+3)(2m_0+3k_0+2)(2m_0+3k_0+1)}\right]\frac{|3
\beta|}{(3!)^22^3}\\ &\approx & 27 \beta k_0 \approx 1 \Rightarrow
k_0 \approx {1\over 27 \beta} \qquad ,
\end{eqnarray*}
and the error assumed by neglecting higher-order terms is bounded
by ${\rm exp}[-{1\over 27 \beta}]$.

Writing the partition function in the form
\begin{equation}
{{\cal Z}[\beta] \over e^{2 \beta \over 3} \sqrt{4 \beta/ 3} } =
\sum_{m=0}^\infty\sum_{k=0}^\infty \int \overbrace{ {(-\beta)
\varphi^4 \over (4!)} \times \cdots \times {(-\beta) \varphi^4
\over 4!} \over m!} \cdot \overbrace{ { (-\sqrt{3 \beta})
\varphi^3 \over 3!}\times \cdots \times { (-\sqrt{3 \beta})
\varphi^3 \over 3!} \over (2k)!} e^{-{1\over 2} \varphi^2 } d
\varphi \label{eq:feyn1}
\end{equation}
one sees that the Feynman rules encompass one tetra-valent vertex
and one trivalent vertex that are proportional respectively to $(-
\beta)$ and $-(\sqrt{3 \beta})$. Four-leg vertices come from
${(-\beta) \varphi^4 \over (4!)}$ in the integrand of
(\ref{eq:feyn1}); three-leg vertices are due to ${ (-\sqrt{3
\beta}) \varphi^3 \over 3!}$ terms in (\ref{eq:feyn1}) and only
contribute in pairs. Comparison with the $\bar{\hbar}$-expansion
(\ref{eq:ass1}) shows that pictures of the $k=0$ terms, collected
in the first two blocks of the first row, are provided by the
diagrams shown in Table 1. Diagrams with one tetra-valent and two
three-valent vertices, $k=m=1$, shown in Table 2, provide the
second block in the first row: $-{ 3 \beta^2 \over 2^6}-{ 3
\beta^2 \over 3 \cdot 2^5 }- { 3 \beta^2 \over 2^4}-{ 3 \beta^2
\over  2^2}- { 3 \beta^2 \over 2^3} -{ 3 \beta^2 \over 3\cdot 2^2}
\equiv - {3\cdot 5 \cdot 7 \over 2^6} \beta^2$.

\begin{table}[h]
\begin{center}
\begin{tabular}{cccccc} \\ \hline
\textit{ Diagram} &  & \textit{ Weight} &
\textit{ Diagram} &  & \textit{ Weight}\\
\hline \\
\parbox{2cm}{\epsfig{file=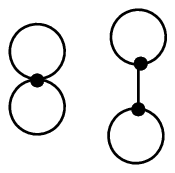,width=1cm}}
&  $\rightarrow$ &  $ {\displaystyle -{ 3 \beta^2 \over 2^6}} $ &
\parbox{2cm}{\epsfig{file=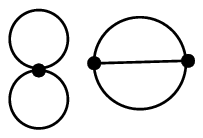,width=1cm}}
&  $\rightarrow$ &  $ {\displaystyle -{ 3 \beta^2 \over 3 \cdot 2^5 }} $ \\  & & & & & \\
\parbox{2cm}{\epsfig{file=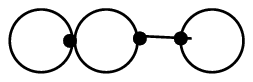,width=1.5cm}}
&  $\rightarrow$ &  $ {\displaystyle - { 3 \beta^2 \over 2^4}}$ &
\parbox{2cm}{\epsfig{file=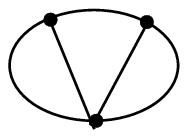,width=0.8cm}}
&  $\rightarrow$ & $  {\displaystyle -{ 3 \beta^2 \over  2^2}} $ \\ & & & & & \\
\parbox{2cm}{\epsfig{file=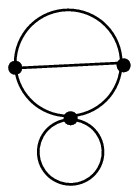,width=0.7cm}}
&  $\rightarrow$ &  $ {\displaystyle - { 3 \beta^2 \over 2^3}} $ &
\parbox{2cm}{\epsfig{file=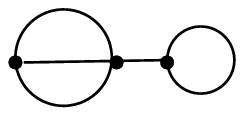,width=1.2cm}}
&  $\rightarrow$ &  $ {\displaystyle -{ 3  \beta^2 \over 3\cdot 2^2}} $ \\
 \hline
\end{tabular} \qquad ,
\end{center}
\end{table}

\noindent In Table 3 only diagrams with tri-valent vertices,
$m=0$, are displayed:
\begin{table}[h]
\begin{center}
\begin{tabular}{cccccc} \\ \hline
\textit{ Vacuum graph} &  & \textit{ Weight} &
\textit{ Vacuum graph} &  & \textit{ Weight}  \\
\hline \\
\parbox{2cm}{\epsfig{file=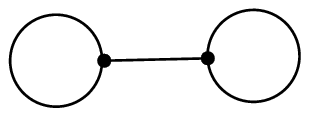,width=1cm}}
&  $\rightarrow$ & $  {\displaystyle 3 \beta \over \displaystyle
2^3}$ &
\parbox{2cm}{\epsfig{file=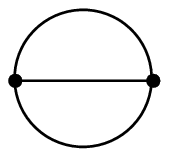,width=0.5cm}}
& $\rightarrow$ & $ {\displaystyle 3 \beta \over \displaystyle
3\cdot 2^2 } $ \\ & & & & & \\
\parbox{2cm}{\epsfig{file=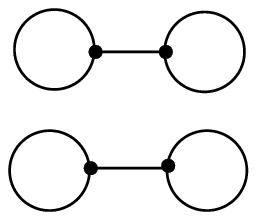,width=1cm}}
& $\rightarrow$ & ${\displaystyle 3^2 \beta^2 \over \displaystyle
2^7}$ &
\parbox{2cm}{\epsfig{file=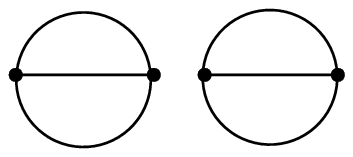,width=1cm}}
&  $\rightarrow$ &  $ {\displaystyle 3^2 \beta^2 \over
\displaystyle 3
\cdot 2^5} $\\ & & & & & \\
\parbox{2cm}{\epsfig{file=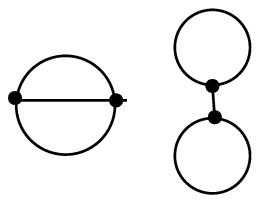,width=1cm}}
&  $\rightarrow$ &  $ { \displaystyle 3^2 \beta^2 \over
\displaystyle 3^2 \cdot 2^5} $ &
\parbox{2cm}{\epsfig{file=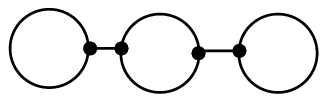,width=1.2cm}} &
 $\rightarrow$ &  $ {\displaystyle 3^2 \beta^2
\over\displaystyle
2^5} $ \\ & & & & & \\
\parbox{2cm}{\epsfig{file=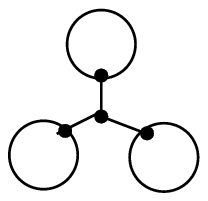,width=1cm}}
& $\rightarrow$ &  $ {\displaystyle 3^2 \beta^2 \over
\displaystyle 3 \cdot 2^4} $ &
\parbox{2cm}{\epsfig{file=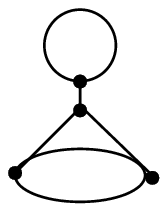,width=0.7cm}}
& $\rightarrow$ &  $ {\displaystyle 3^2 \beta^2 \over \displaystyle 2^4} $ \\ & & & & & \\
\parbox{2cm}{\epsfig{file=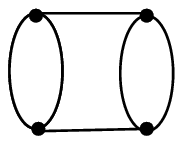,width=0.7cm}}
&  $\rightarrow$ & $ { \displaystyle 3^2 \beta^2 \over
\displaystyle 2^5} $ &
\parbox{2cm}{\epsfig{file=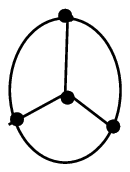,width=0.5cm}}
&  $\rightarrow$ &  $ {\displaystyle 3^2 \beta^2 \over \displaystyle 3 \cdot 2} $ \\
 \hline
\end{tabular}
\end{center}
\end{table}

\noindent Diagrams with two tri-valent vertices contribute: $ { 3
\beta \over 2^3} + { 3 \beta \over 3\cdot 2^2 } \equiv {5\over
2^3} \beta $, whereas the contribution of diagrams with four
trivalent vertices is: ${3^2 \beta^2 \over 2^7} + {  3^2 \beta^2
\over 3 \cdot 2^5} + { 3^2 \beta^2 \over 2^5} + { 3^2 \beta^2
\over 3 \cdot 2^4}
 + {  3^2 \beta^2 \over   2^4} + {3^2 \beta^2 \over   2^5} + {  3^2 \beta^2
\over  3 \cdot 2} \equiv {5\cdot 7\cdot 11 \over 2^7} \beta^2$.

\section{Supersymmetric  WKB approximation}

The semiclassical regime is characterized by the inequality:
\[
\bar{\hbar}\left|{d^2W\over dx^2}(x)\right|<< \left|{dW\over
dx}(x)\right|^2=|\Omega(x)\cdot\Omega(x)|^2 \qquad , \qquad
\Omega(x)=\sqrt{2V(x)} \hspace{2.7cm}\quad . \label{ineq:sc}
\]
Thus,
\[
{\bar{\lambda}\over 2\pi}={\bar{\hbar}\over\sqrt{2V(x)}}
<<\left|\frac{2V(x)}{dV/dx}\right|
\]
is satisfied in the limit of short wave lengths. To obtain the WKB
eigen-functions of the SUSY Hamiltonian in, e.g. , the subspace
for which the zero Fermi number is zero - because supersymmetry
WKB eigenfunctions of non-zero energy in the Fermi sector are
given automatically - one starts from the
Wentzel-Krammers-Brillouin ansatz in the classically forbidden
region $E<V(x)$:
\begin{equation}
\Psi^E(x,t)=A(x)\cdot{\rm exp}[-\frac{ W_E(x)}{\bar{\hbar}}]\cdot
e^{i\frac{Et}{\bar{\hbar}}} \label{eq:wkb1}\qquad .
\end{equation}
The Schrodinger equation for $V_+(x)$ becomes
\begin{eqnarray}
&&\bar{\hbar}^2\left({d^2\ln A\over dx^2}(x)+{d\ln A\over dx}(x)
\cdot{d\ln A\over
dx}(x)\right)-\nonumber\\&-&\bar{\hbar}\left({d^2W_E\over
dx^2}(x)+2\ln A(x)\cdot{dW_E\over dx}(x)-{d\Omega\over
dx}(x)\right)+W_E(x)\cdot W_E(x)-\Omega(x)\cdot\Omega(x)+2E=0
\quad . \label{eq:wsch}
\end{eqnarray}
with three terms of respectively order 2,1, and 0 in
$\bar{\hbar}$. The usual WKB strategy starts by solving the
equation (\ref{eq:wsch}) for the $\bar{\hbar}$-independent terms
to find:
\[
W_E(x)=\int \, dx \, \sqrt{\Omega(x)\cdot\Omega(x)-2E} \qquad ,
\]
with the novelty with respect to the non SUSY case that the
turning points are those corresponding to $V(x)$, rather than
those set by the effective potential $V_+(x)$. The second step is
to plug this solution into the equation for the terms proportional
to $\bar{\hbar}$:
\[
\frac{d\ln A}{dx}(x)={1\over
2}\left[\frac{1}{\sqrt{\Omega(x)\Omega(x)-2E}}-\frac{\Omega(x)}{\Omega(x)\Omega(x)-2E}\right]
\cdot {d\Omega\over dx}(x) \qquad .
\]
Integration of this equation provides the SUSY WKB wave functions:
\begin{equation}
A(x)\propto \frac{1}{(\Omega(x)\Omega(x)-2E)^{{1\over 4}}}\cdot
\left(|\Omega|(x)+\sqrt{\Omega(x)\Omega(x)-2E}\right)^{{|\over 2}}
\label{eq:fcwkb} \qquad .
\end{equation}
Note the other difference: in the non-SUSY case the numerator of
this expression is $1$. In the classical allowed regions,
$2E>\Omega^2(x)$, however, the WKB ansatz reads,
\begin{equation}
\psi^E(x)=A(x){\rm exp}[-i\frac{|W(x)|}{\bar{\hbar}}]
\label{eq:wkb2} \qquad ,
\end{equation}
and one obtains:
\[
A(x)={1\over \left[2E-\Omega^2(x)\right]^{{1\over 4}}}\cdot{\rm
exp}\left[{i\over 2}{\rm arcsin}\frac{\Omega(x)}{\sqrt{2E}}\right]
\qquad .
\]
To match the WKB wave functions (\ref{eq:wkb1}) and
(\ref{eq:wkb2}) analytically at the classical turning points
$x=a\,< \,x=b$, such that $\Omega^2(a)=2E=\Omega^2(b)$, the
following supersymmetric quantization rule is required:
\begin{equation}
\int_a^b \, dx \, \left[2E-\Omega^2(x)\right]^{{1\over
2}}=n\pi\bar{\hbar} \qquad \qquad n\in{\mathbb Z}^+ \qquad .
\label{eq:swkb}
\end{equation}
The appearance of the numerator in (\ref{eq:fcwkb}) is magic:
firstly, because this term modifies the process of analytic
continuation necessary to match the exponential and periodic WKB
wave functions at the turning points in such a way that the
${\pi\over 2}\bar{\hbar}$ term that appears in the non SUSY
version of (\ref{eq:swkb}) does not enter the SUSY case. To obtain
the WKB wave function in the classically allowed region
\[
\psi^E(x)=e^{-\sqrt{{E\over
2}}}\cdot\frac{\sqrt{\sqrt{2E-\Omega^2}+i\Omega}}{(2E-\Omega^2)^{1\over
4}}\cdot\left\{C_1e^{{i\over\bar{\hbar}}\int_b^x\, dx'\,
\sqrt{2E-\Omega^2(x')}}+C_2e^{-{i\over\bar{\hbar}}\int_b^x\, dx'\,
\sqrt{2E-\Omega^2(x')}}\right\}
\]
from the WKB wave functions in the forbidden regions
\[
\psi^E(x)=C\frac{\sqrt{\sqrt{\Omega^2-2E}+|\Omega|}}{(\Omega^2-2E)^{1\over
4}}\cdot e^{-{1\over\bar{\hbar}}\int_x^a\, dx'\,
\sqrt{\Omega^2(x')-2E}} \hspace{0.15cm} ; \hspace{0.15cm}
\psi^E(x)=C'\frac{\sqrt{\sqrt{\Omega^2-2E}+|\Omega|}}{(\Omega^2-2E)^{1\over
4}}\cdot e^{-{1\over\bar{\hbar}}\int_b^x\, dx'\,
\sqrt{\Omega^2(x')-2E}}
\]
one chooses paths in the $x$-complex plane that goes around the
turning points $a$ and $b$ at great distance, either in the upper
or the lower half-planes. Unlike to the non-SUSY case, there is no
$e^{-i{\pi\over 4}}$ factor left and two wave functions are
obtained in the classically allowed region, one from the left and
the other from the right:
\[
\psi^E(x)=C\frac{\sqrt{\sqrt{2E-\Omega^2(x)}+i\Omega(x)}}{(2E-\Omega^2(x))^{1\over
4}}\cdot{\rm cos}[{1\over\bar{\hbar}}\int_b^x\, dx'\,
\sqrt{2E-\Omega^2(x')}]
\]
\[
\psi^E(x)=C'\frac{\sqrt{\sqrt{2E-\Omega^2(x)}+i\Omega(x)}}{(2E-\Omega^2(x))^{1\over
4}}\cdot{\rm cos}[{1\over\bar{\hbar}}\int_x^a\, dx'\,
\sqrt{2E-\Omega^2(x')}] \qquad .
\]
These expressions are identical if and only if (\ref{eq:swkb})
holds. Secondly, $E=0$ is a solution of (\ref{eq:swkb}) for $n=0$,
whereas (\ref{eq:wkb1}) becomes the exponential of the
superpotential: the exact ground state is a SUSY WKB wave function
!

\subsection{WKB analysis of the single well}

We shall consider as examples non-harmonic oscillators of fourth
order to avoid hyperelliptic integrals and deal with (slightly!)
manageable expressions. In the case of a single well with
potential energy $V(x)={\lambda\over 2}(x^2+{k\over\lambda})^2$ we
have, using non-dimensional variables:
\[
V(x)={1\over 2}(x^4+2x^2+1) \qquad , \qquad \Omega(x)=x^2+1 \qquad
, \qquad W(x)={x^3\over 3}+x \qquad .
\]
The turning points are the real roots of the quartic equation:
\begin{equation}
x^4+2x^2-a=0 \qquad , \qquad a=2E-1 \qquad , \qquad
x_\pm=\pm\sqrt{-1+\sqrt{1+a}} \label{eq:quar1} \qquad .
\end{equation}
The supersymmetric quantization rule is therefore:
\begin{equation}
I(E;x_-,x_+)=\int_{x_-}^{x_+} \sqrt{a-x^4-2x^2}\,
dx=n\pi\bar{\hbar} \label{eq:qwkb1} \qquad .
\end{equation}
Denoting $A_\pm=\pm1+\sqrt{1+a}$, the definite integral in
(\ref{eq:qwkb1}) reads:
\begin{equation}
I(E;x_-,x_+)=\frac{4}{3} \sqrt{A_-} \left( \sqrt{a+1} \,
\mathbb{K}\left( \frac{-A_-}{A_+}\right) -\mathbb{E}\left(
\frac{-A_-}{A_+}\right)\right)\qquad ,
\end{equation}
where $\mathbb{K}(k^2)$ and $\mathbb{E}(k^2)$ are respectively the
complete elliptic integrals of first and second type. This result
is shown in Figure 8.

\noindent\begin{figure}[htbp] \centerline{
 \epsfig{file=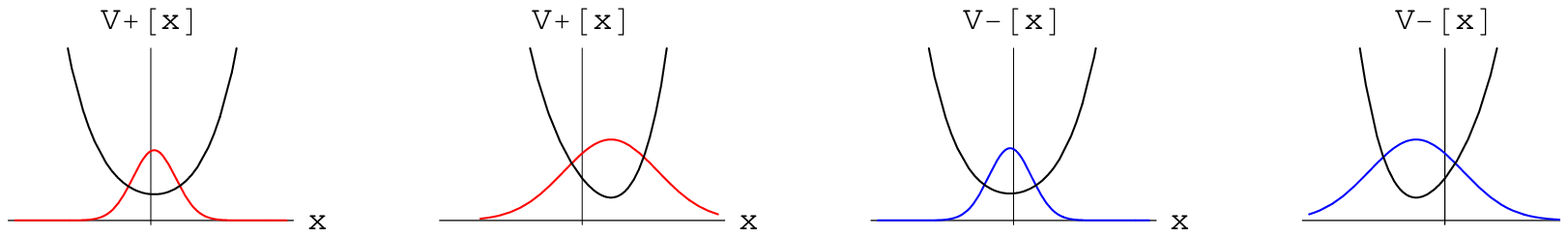,height=2cm}}
 \caption{Graphics of $V^\pm(x)$ for
 $\bar{\hbar}=0.1$ -(1),(3)- and  $\bar{\hbar}=1$ -(2),(4)-. }
\end{figure}

\noindent\begin{figure}[htbp] \centerline{
 \epsfig{file=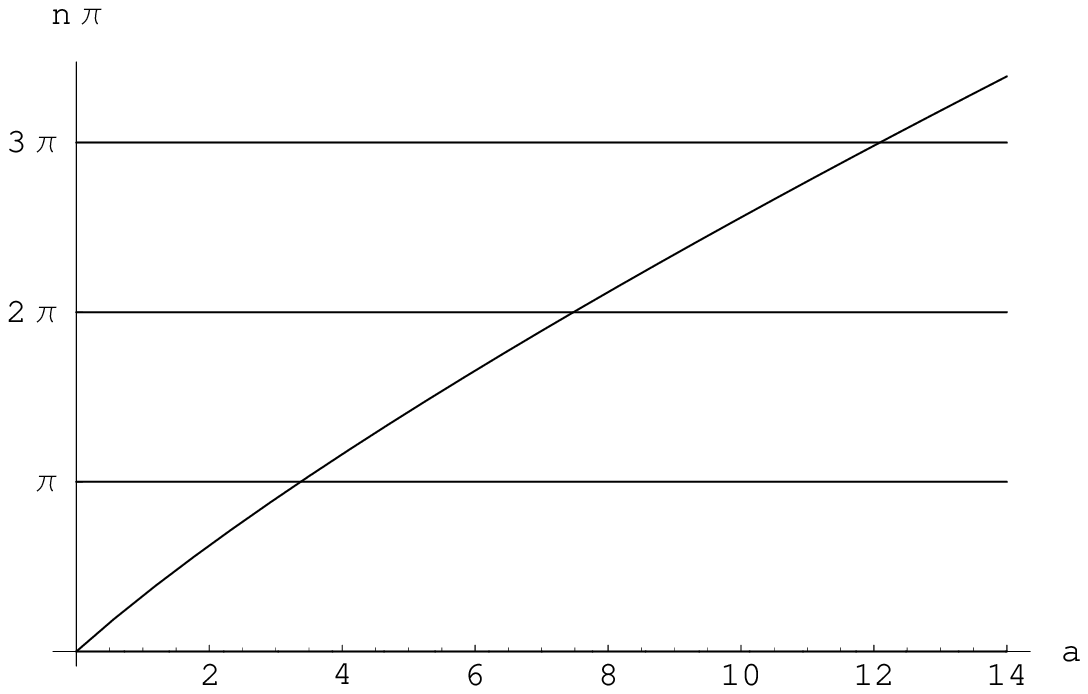,height=4.5cm}\hspace{0.5cm} \epsfig{file=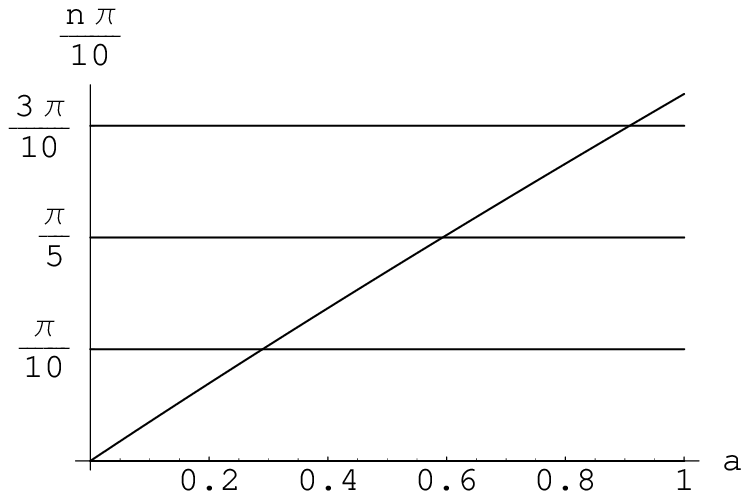,height=4cm}}
 \caption{Mathematica plot of $I(E;x_-,x_+)$ as a function of $a$ and intersection with $n\pi\bar{\hbar}$ for low $n$ and
 $\bar{\hbar}=1$ (left) and  $\bar{\hbar}=0.1$ (right)}
\end{figure}

The first three (double, see Figure 7) eigenvalues for
$\bar{\hbar}=1$ and $\bar{\hbar}=0.1$ are: $E_1=2.18674
{k^2\over\lambda}$, $E_2=4.23942\,{k^2\over\lambda}$,
$E_3=6.5444\,{k^2\over\lambda}$, and
$E_1=0.64500\,{k^2\over\lambda}$, $E_2=0.78289\,
{k^2\over\lambda}$, $E_3=0.95403\,{k^2\over\lambda}$,
respectively.


\subsection{WKB analysis of the double well}

For a non-harmonic oscillator of fourth order and a double well
things are even more difficult. The potential energy is
$V(x)={\lambda\over 2}(x^2-{k\over\lambda})^2$, such that in
non-dimensional variables we have:
\[
V(x)={1\over 2}(x^4-2x^2+1) \qquad , \qquad \Omega(x)=x^2-1 \qquad
, \qquad W(x)={x^3\over 3}-x \qquad .
\]
The turning points are the real solutions of the quartic equation:
\begin{equation}
x^4-2x^2-a=0 \qquad , \qquad x_{-\,\mp}=
-\sqrt{1\mp\sqrt{1+a}}\qquad , \qquad
 x_{+\,\mp}=\sqrt{1\mp\sqrt{1+a}} \qquad .\label{eq:quar2}
\end{equation}
For $a>0$ there are only two real roots and the supersymmetric
quantization rule reads:
\begin{equation}
I(E;x_{-+},x_{++})=\int_{x_{-+}}^{x_{++}} \sqrt{a-x^4+2x^2}\,
dx=n\pi\bar{\hbar} \label{eq:qwkb1} \qquad .
\end{equation}
The computation of $I(E;x_{-+},x_{++})$ is qualitatively identical
to the previous case and results are shown in Fig. 10(left).

If $-1<a<0$ things are more difficult: there are four turning
points, four real roots, and the quantization rule splits into two
equations:
\begin{equation}
I(E;x_{--},x_{+-})=\int_{x_{--}}^{x_{+-}} \sqrt{a-x^4+2x^2}\,
dx=n\pi\bar{\hbar}=I(E;x_{-+},x_{++})=\int_{x_{-+}}^{x_{++}}
\sqrt{a-x^4+2x^2}\, dx \quad . \label{eq:qwkb12}
\end{equation}
The definite integrals in (\ref{eq:qwkb12}) now read:
\begin{eqnarray}
I(E;x_{--},x_{+-})=I(E;x_{-+},x_{++})=&&\frac{2a}{3 A_+
\sqrt{A_-}} \left[ \sqrt{a+1} \, \left( \mathbb{K}\left(
\frac{-A_+}{A_-}\right)-\mathbb{F}\left(\arcsin\sqrt{\frac{-A_-}{A_+}}
,\frac{-A_+}{A_-} \right)\right)\right. \nonumber\\ && \left.
-\mathbb{E}\left(
\frac{-A_-}{A_+}\right)+\mathbb{E}\left(\arcsin\sqrt{\frac{-A_-}{A_+}}
,\frac{-A_+}{A_-} \right)\right] \qquad . \label{eq:elip}
\end{eqnarray}
Note that incomplete elliptic integrals of the first, ${\mathbb
F}(u,m)$, and second, ${\mathbb K}(u,m)$, type also enter. In any
case, it is possible to plot these functions of $a$ and find the
intersection points determining the spectrum.

\noindent\begin{figure}[htbp] \centerline{
 \epsfig{file=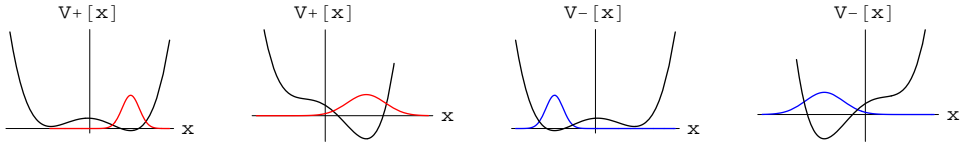,height=2cm}}
 \caption{\small Graphics of $V^{\pm}(x)$ for
 $\bar{\hbar}=0.1$ -(1),(3)- and  $\bar{\hbar}=1$ -(2),(4)-.}
\end{figure}

\noindent\begin{figure}[htbp] \centerline{
 \epsfig{file=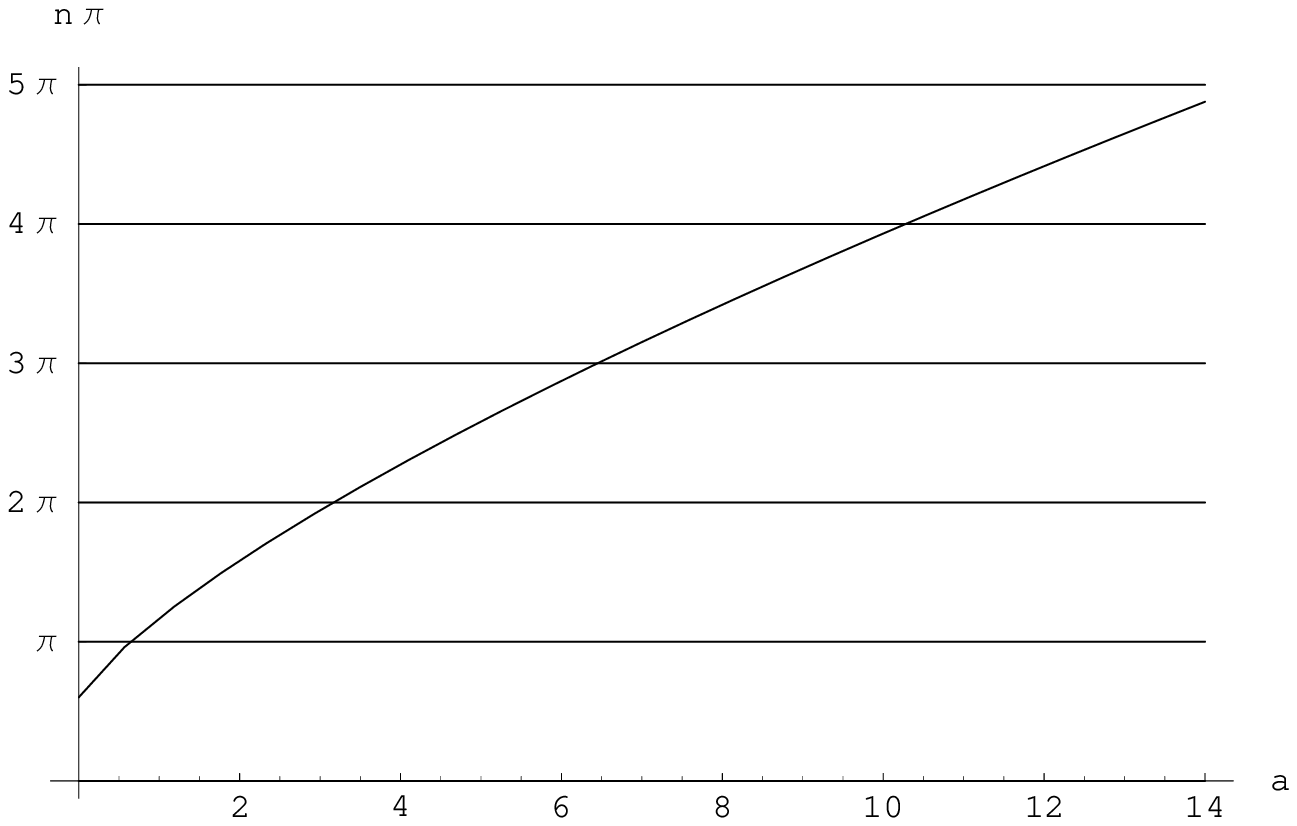,height=4cm}\hspace{0.5cm} \epsfig{file=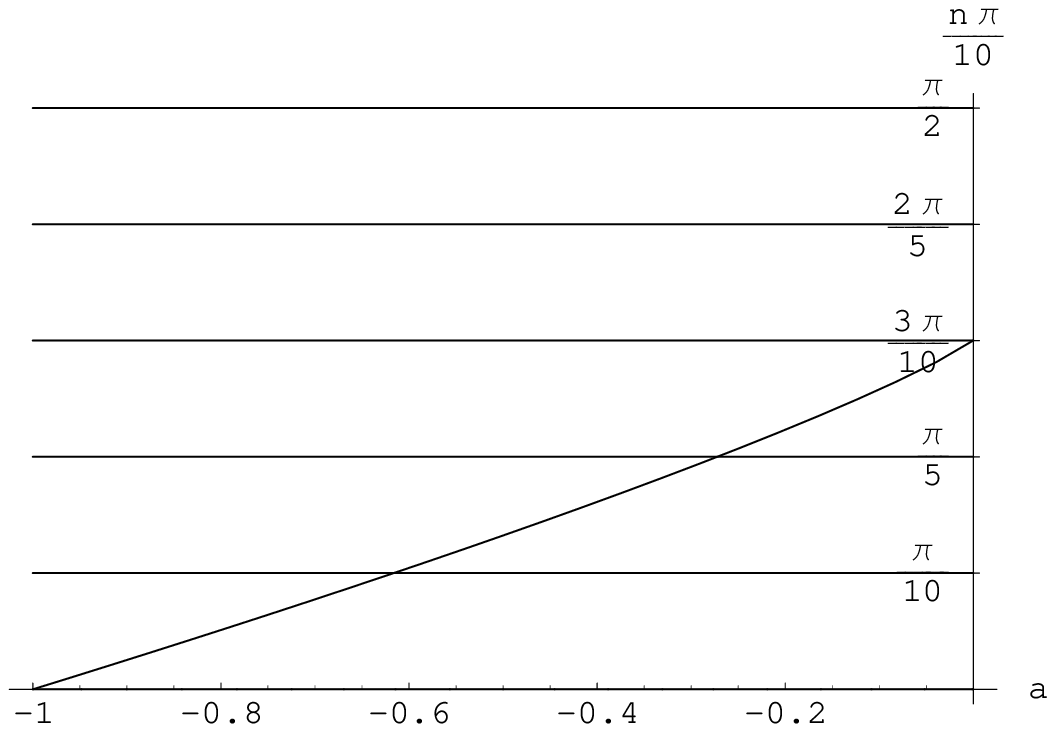,height=3.7cm}}
 \caption{\small Mathematica plots of $I(E;x_{-+},x_{++})$ for
 $\bar{\hbar}=1$ and $a>0$ (left) and $I(E;x_{--},x_{+-})$ for
 $\bar\hbar=0.1$ and $-1<a<0$ (right) as function of $a$. The
 intersection points with $n\pi\bar\hbar$ giving the eigenvalues
 are also shown.}
\end{figure}

The first three eigenvalues for $a>0$ and $\bar\hbar=1$ are: $
E_1=0.82272 \,{k^2\over\lambda}$,
$E_2=2.08330\,{k^2\over\lambda}$,
$E_3=5.63830\,{k^2\over\lambda}$.

In the case of $-1<a<0$ eigenvalues only exist if
$\bar\hbar<0.95$. Application of rule (\ref{eq:qwkb12}) for the
turning points on the left gives: $E_1=0.19183
\,{k^2\over\lambda}$, $E_2=0.36384\,{k^2\over\lambda}$,
$E_3=0.49993\,{k^2\over\lambda}$.

Because of formula (\ref{eq:elip}) the choice of pair of turning
points is irrelevant; $E_1$, $E_2$, $E_3$, etcetera, are
eigenvalues of the Schr\"odinger equation for both $V_+(x)$ and
$V_-$.

\section{Outlook}

The next step is to study physical systems of two degrees of
freedom. It is tempting to start by discussing problems of this
type in Hamilton separable systems. Following the works
\cite{Ioffe}-\cite{Ioffe2} on supersymmetric quantum mechanics in
more than one dimensions, the general structure of supersymmetric
classical and quantum Liouville systems has been described in
References \cite{AoP} and \cite{JPA}. An important example of this
kind of systems is the supersymmetric classical and quantum
hydrogen atom respectively analyzed by Heumann \cite{Heumann} and
Kirchberg et al \cite{KLPW}. It seems also plausible to address
similar issues in non-separable but integrable systems as those
proposed in \cite{Ioffe3}.

\section{Acknowledgements}

We thank M. Ioffe for enlightening conversations on supersymmetric
quantum mechanics and a critical reading of the manuscript. This
work has been partially supported by the Spanish MEC grant
BFM2003-00936 and the project VA013C05 of the Junta de Castilla y
Leon.

\end{document}